\documentstyle[12pt,aps,preprint,axocolor]{revtex}
\tightenlines

\newcommand{\be}{\begin{equation}}
\newcommand{\ee}{\end{equation}}
\newcommand{\ba}{\begin{array}{c}}
\newcommand{\ea}{\end{array}}
\newcommand{\bqa}{\begin{eqnarray}}
\newcommand{\eqa}{\end{eqnarray}}
\newcommand{\tm}{\tilde m}
\newcommand{\x}{\phi_V}
\begin{document}
\draft
\title{On the Long Distance Contribution to the $B_s\to \gamma\gamma$ Decay
in the Effective Lagrangian Approach}
\author{Wei Liu, Bin Zhang  and  Hanqing Zheng}
\address{Department of Physics, Peking University, Beijing 100871,
People's Republic of China}
\maketitle

\begin{abstract}We re-estimate the decay branching ratio of $B_s\rightarrow 2\gamma$
through  $D_s^+D_s^-$ and $D_s^{*+}D_s^{*-}$ intermediate states,
in the   effective Lagrangian approach. We find that the
branching ratio does not exceed a few times $10^{-7}$, contrary to
the result recently claimed in the literature.
\end{abstract}

\begin{figure}[htb]
\input{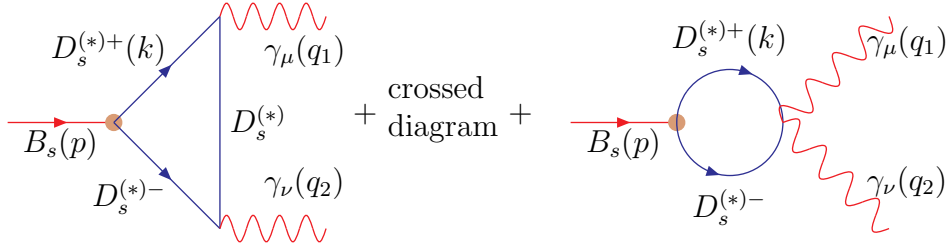}
\vskip 0.5 cm
   \caption{\em One-loop contribution to the
        $B_s \to \gamma\gamma$
       amplitude due to $D_s$ ($D_s^*$)  alone.
       }
 \label{fig1}
\end{figure}

\vspace{1cm}
The $B_s \rightarrow \gamma\gamma$ decay  is an intereting process
which has been investigated by several
groups~\cite{LLY,Sim_Wyl,Kalin} within the context of  pQCD. The
branching ratio was found to be ${\cal BR}(B_s \rightarrow
\gamma\gamma)\sim 3.8 \times 10^{-7}$ for $m_t\simeq 175GeV$, to be
compared with the   present experimental upper limit ${\cal BR}(B_s
\rightarrow \gamma\gamma) < 1.48 \times 10^{-4}$~\cite{L3}. It was
also pointed out that the branching ratio can be substantially
enhanced  in some  extensions of the standard model such as a
generic 2-Higgs doublet model~\cite{2Higgs}. Provide that the short
distance contribution is not large~\cite{CLY,2Higgs,SRC_Yao}, one
may then hope  the observation of such channel in future
experiments as a signal of  physics beyond  standard model.
However, as have been emphasized by Choudhury  and  Ellis~\cite{CE},
a careful analysis to the long distance effects has to be made
before one could use this channel as a probe to the physics beyond
the standard model.  Choudhury  and  Ellis have considered the
contributions due to intermediate $D_s$ and $D^*_s$  states via the
diagrams
 including loops of $D_s$ mesons alone,
loops of $D_s^*$ mesons alone~(see fig.~\ref{fig1}), and diagrams
involving radiative $D_s^*
\rightarrow D_s + \gamma$ transitions. Based upon  an estimation
on the absorptive part of the amplitude they find that the
contribution due to $D_s$ alone is,
\be\label{ds}
 Br (B_s \rightarrow 2 D_s \rightarrow
 \gamma \gamma) \sim 2.9 \times 10^{-8}\ ,
\ee
which is insignificant as comparing with the pQCD short distance
contribution. Diagrams involving the $D_s^*\to D_s\gamma$
transition are again found to be small. However they find that the
contribution due to $D_s^*$  is much larger,
\be\label{dstar}
 Br (B_s \rightarrow 2 D_s^*\rightarrow
 \gamma \gamma) \sim 6.5 \times 10^{-6}\ ,
\ee
 which, if correct, is of  great phenomenological interets as it will be easily seen
in future hadron colliders.
Because of its importance it is worthwhile to give a careful
re-analysis to such a process.

%%%

We in the following use the same method as was adopted in
ref.~\cite{CE}, i.e.,  a diagramatic analysis of fig.~\ref{fig1}
based on an effective Lagrangian approach. Beside those having been
done in ref.~\cite{CE} we also estimate the real part contribution
of the rescattering amplitude using dispersion relations. The
Feynman rules we use for the calculation are derived from the
dimension 4, $U_{em}(1)$ gauge invariant effective Lagrangian which
are bilinear in $D_s$ fields:
\bqa\label{Lag}
{\cal L}_{eff}&=&\nabla_\mu D^+\nabla^\mu D^--{1\over
2}\left(\nabla_\mu D^-_\nu-\nabla_\nu D^-_\mu\right
)^+\left(\nabla^\mu D^{-\nu}-\nabla^\nu D^{-\mu}\right
)+ie\phi_VD_\mu^+F^{\mu\nu}D_\nu^-\ ,\\
\nabla_\mu &=&\partial_\mu-ieA_\mu \ ,\nonumber
\eqa
where $\phi_V$  is an unknown parameter not
constrained by gauge invariance. We have verified that the choice
made in ref.~\cite{CE} corresponds to $\phi_V=0$. In the case of
elementary $W^+W^-\gamma$ interaction $\phi_V=1$ and in the case of
effective $\rho^+\rho^-\gamma$ or $K^{*+}K^{*-}\gamma$ interactions
the parameter $\phi_V$ can be estimated from the  extended
Nambu--Jona-Lasinio model and is found to be slightly less than
1~\cite{prades} (Our definition of $\phi_V$ is two times larger
than Ref.~\cite{prades}). Since we did not find convincing reason
to support taking $\phi_V=0$, we treat $\phi_V$ as a free parameter
ranging from $0$ to $1$.

 The absorptive part contribution of fig.~\ref{fig1} can be
written as,
\be\label{abs}
{\rm Im}A=\Sigma A_0^{\lambda_1\lambda_2}\beta T_s^{\lambda_1\lambda_2,\lambda_3\lambda_4}\ ,
\ee
where $\beta=\sqrt{1-4m^2/s}$ is the kinematic factor,   the  sum
is over the helicity indices $\lambda_1,
\lambda_2$  of the intermediate states
(for $D^{*+}$ and $D^{*-}$ scattering) and $\lambda_3$, $\lambda_4$  are
the helicity indices of the outgoing photons;
 $(\lambda_1,\lambda_2)$ can take $(+,+)$, $(-,-)$ and $(0,0)$,
and $(\lambda_3,\lambda_4)$ can take
$(+,+)$ and $(-,-)$. In above $A_0$ is the decay vertex obtained from factorization method,
\be\label{bdd}
A_0=\langle D_s^+ (k) \; D_s^-(q) | H_{wk} | B_s(p) \rangle
               =  -i g f \; \left[ (p^2 - k^2) f^+(q^2)
               + q^2 f^-(q^2) \right]\ ,
\ee
for $B_s\to D_s^+D_s^-$, and,
\bqa\label{bdd1}
A_0&=&\langle D_s^{\ast +} (k) \; D_s^{\ast -} (q) | H_{wk} | B_s(p) \rangle\nonumber\\
        &=&  g f_\ast \epsilon^{\ast \mu}_{+} \epsilon^{\ast \nu}_{-}
\left[ {\cal V}(q^2) \epsilon^{\nu \mu \alpha \beta} p_\alpha k_\beta
           + i \{ {\cal A}_1(q^2) p^2 g^{\mu \nu}
                                - 2 {\cal A}_2(q^2) p^\mu p^\nu \}\right]\ ,
\eqa
for $B_s\to D_s^{*+} D_s^{*-}$. In above equations, $g={G_F\over
\sqrt{2}}V_{cb}V^*_{cs}$, $f,f^*\simeq 200MeV$, $f_{\pm},{\cal
V}\simeq 0.6$ and ${\cal A}_1,{\cal A}_2\simeq 0.25$, according to Ref.~\cite{CE}.  
The matrix element for the decay 
$B_s \rightarrow \gamma(q_{1\mu}) \gamma(q_{2\nu})$ 
are parametrized as, 
\bqa
    A    & = &  \alpha g 
\left( R_1 S_{\mu \nu} + i R_2 P_{\mu \nu} 
                          \right)\ , \nonumber \\
    S_{\mu \nu} & \equiv & 
                  q_{1\nu} q_{2 \mu} - q_1 \cdot q_2 g_{\mu \nu}\ ,\nonumber \\
  P_{\mu \nu} & \equiv &  \epsilon_{\mu \nu \alpha \beta}
                        q_1^{\alpha} q_{2}^{\beta}\ , 
\eqa
where $R_1$ and $R_2$ are the yet-to-be-determined hadronic matrix
elements, and 
the partial width is then given by,
\be
	\Gamma(B_s \rightarrow \gamma \gamma) = \frac{(\alpha g)^2}{64 \pi} m_B^3\left( |R_1|^2 + |R_2|^2 \right)\ .
\ee
Since $\Gamma (B_s) \simeq 5 \times 10^{-10} MeV$, we have, 
for $|V_{cb}| = 0.04$, 
\be
{
   Br (B_s \rightarrow \gamma \gamma)  \simeq
          10^{-7} 
               \; \left(  | \frac{R_1}{100 MeV} |^2
                       + | \frac{R_2}{100 MeV} |^2
                  \right)
}\ .
\ee

In re-calculating  the processes, for the $D_s^+D_s^-$ intermediate 
state we confirm the result of
Ref.\cite{CE}, Eq.~(\ref{ds}). For the $D_s^{*+}D_s^{*-}$
intermediate state, we have,
\bqa
{\rm Im}R_1 &=& {f_*\over 8 {\tm}^2}{\cal A}_1  \{[(1-4\tm)+\x
(4\tm)-\x^2 ]\lambda_{22} \nonumber\\ &&+\tm
[(1-12\tm+48\tm^2)+\x(-2-8\tm)+\x^2(1-4\tm)]\log({1-\beta\over
1+\beta})\}\nonumber\\ && -{f_*\over 8 {\tm}^2}{\cal
A}_2\{[(1-5\tm)+\x (2\tm)+\x^2(-1+3\tm)]\lambda_{22}
\nonumber \\ &&+\tm
[(1-10\tm+32\tm^2)+\x(-2+4\tm)+\x^2(1-2\tm)]\log({1+\beta\over
1-\beta})\}\ ,\nonumber\\ {\rm Im}R_2&=&{f_*{\cal V}\over 16\tm}
\{[(-1+12\tm)+\x (2-8\tm)+\x^2(-1-4\tm)]\lambda_{22}\nonumber\\&& +
[(4\tm-32\tm^2)+\x(8\tm-32\tm^2)+\x^2(4\tm)]\log({1+\beta\over
1-\beta})\}\ .
\eqa
The calculation made in ref.~\cite{CE} corresponding to  taking
$\phi_V=0$ in above equations. We  find a disagreement on the first
term inside the square bracket in the coefficient of ${\cal A}_2$.
As a consequence, the  CP conserving branching ratio becomes much
smaller, ${\cal BR}(B_s\rightarrow 2\gamma)\simeq 0.85\times
10^{-8}$. Meanwhile we reconfirm the numerical result on the
$CP$ violating branching ratio of ref.~\cite{CE}. The maximal value
of ${\cal BR}(B_s\rightarrow 2\gamma)$ comes from the case
$\phi_V=1$. In this situation, our calculation is equivalent to
taking the $W^+W^-\gamma$ vertics (propagators are in the unitary
gauge). Hence we list some intermediate steps in the calculation
here since it may be useful elsewhere. The $s$ partial--wave
projection of the $2\to 2$ scattering amplitudes are found to be,
\be\label{ds0}
\beta T_s(D_s^+D_s^-\rightarrow 2\gamma)={\alpha_e\over
2s}(1-\beta^2)\ln({1+\beta\over 1-\beta}) \ ,
\ee
and,
\bqa\label{ds1}
\beta T_s^{++}(D_s^{*+}D_s^{*-}\rightarrow 2\gamma)&=& {\alpha_e\over 2s}
(1+\beta)^2\ln\left({1+\beta\over 1-\beta }\right)\ ,
\nonumber\\
\beta T_s^{--}(D_s^{*+}D_s^{*-}\rightarrow 2\gamma)&=& {\alpha_e\over 2s}
(1-\beta)^2\ln\left({1+\beta\over 1-\beta }\right)\ ,
\nonumber\\
\beta T_s^{00}(D_s^{*+}D_s^{*-}\rightarrow 2\gamma)&=&{\alpha_e \over 8m_{D^*}^2}
(1-\beta^2)^2
\ln \left({1+\beta\over 1-\beta }\right)\ .
\eqa
In above the gauge invariant factor
$\epsilon_1\cdot\epsilon_2k_1\cdot k_2-\epsilon_1\cdot
k_2\epsilon_2\cdot k_1$ ($=s/2$) has been taken away, and the photon helicity is taken as (+,+). Once again we
find that the $D_s^{*+}D_s^{*-}$ intermediate state contribution to
the decay branching ratio of $B_s\to \gamma\gamma$ quite small,
though it is comparable in magnitude to the short distance
contributions,
\be\label{d*b}
{\cal BR}_{CP-even} (B_s \to 2 D_s^*\rightarrow 2\gamma )
\sim 1.49 \times 10^{-7}\ ,\,\, {\cal BR}_{CP-odd} (B_s \rightarrow 2 D_s^*\to 2\gamma )
\sim 0.44 \times 10^{-7}\ .
\ee

From the above results we find that the decay branching ratio of
$B_s\rightarrow \gamma\gamma$ will not exceed a few times $10^{-7}$
provide that the real part contribution does not change the above
estimtate in order of magnitude. It is natural to expect that the
real part contribution is comparable in order of magnitude to the
absorptive part. Therefore, if the future experiments reveal a
large branching ratio of $B_s\to 2\gamma$ upto a few times
$10^{-6}$ it is very likely that it is a signal of new physics.

Now we discuss the real part contribution of the process depicted
in fig.~\ref{fig1}. The real part contribution can be calculated using
dispersion relation. When performing the dispersion integral it is
necessary to extract the $\epsilon_1\cdot\epsilon_2k_1\cdot
k_2-\epsilon_1\cdot k_2\epsilon_2\cdot k_1$ part out of the
dispersive integral~\cite{mrs} since it is the remaining part to be
considered as the form-factor.\footnote{For earlier referrences on
the usage of dispersion relation under  such a situation, see
ref.~\cite{mrs,Uy}.} We get,
\be\label{fsi}
{\bf A}= {1\over \pi}
\int^\infty_{4m^2}{{\rm Im}{\bf A}\over s'-s-i\epsilon}ds'\ .
\ee

The $D_s$ loop in
fig.~1 contains no ultra-violet divergence and it is
straightforward to evaluate Eq.~(\ref{fsi}), we find,
\be\label{dsnew}
 {\cal BR} (B_s \rightarrow 2 D_s \rightarrow
 \gamma \gamma) \sim 4.2 \times 10^{-8}\ .
\ee
For the $D_s^*$ contribution the diagram fig.~1 is ultra-violet
divergent and needs to be renormalized in the standard perturbation
calculation. In the dispersive approach the divergence manifests
itself by the fact that the dispersive integral in Eq.~(\ref{fsi})
needs one subtraction. The divergence can also be rescued by
introducing form-factors in the Feynman vertex to regulate the high
energy behaviour of the Feynman diagram. However, it is equivalent
in principle to regulate Eq.~(\ref{fsi}) by using an un-subtracted
dispersion relation with truncated integrand,
\be\label{fsitr}
{\bf A}= {1\over \pi}\int^{\Lambda^2}_{4m^2}{{\rm Im}{\bf A}\over
s'-s-i\epsilon}ds'\ .
\ee
Of course, the ultra-violet divergence emerged from the effective
Lagrangian approach soley indicates that the effective Lagrangian
approach gives a bad high energy behaviour to the rescattering
amplitude. For example, the $T_s^{++,++}$ amplitude of
Eq.~(\ref{ds1}), behaves as $\sim {1\over s}\ln(s)$ as
$s\rightarrow \infty$ and does not satisfy Eq.~(\ref{fsi}).
However, from Regge theory we know that, at high energies, the
s--wave scattering amplitudes behave as $T_s(s)\sim
s^{\alpha_{D_s}-1}$ and $T_s(s)\sim s^{\alpha_{D_s^*}-1}$, for the
$D_s$ and $D_s^*$ exchanges, respectively. The intercept parameters
of the Regge  trajectory are $\alpha_{D_s}\simeq -1.5$ and
$\alpha_{{D_s}^*}\simeq -1$.  It is clear that the dispersion
relation, Eq.~(\ref{fsi}), is convergent. The rapid fall-off 
behaviour of
the Regge amplitude with respect to s  implies that the magnitude
of  the real part contribution obtained in the effective Lagrangian approach and via Eq.~(\ref{fsitr})  be well overestimated. In
fact,  Regge behaviour should already dominate at
$s=M_{B_s}^2$~\cite{hz}. So the result obtained from
Eq.~(\ref{fsitr}) may at best be interpreted as an upper bound.
Since the high energy contribution can not be important to the
dispersive integral in Eq.~(\ref{fsi}),  we take $\Lambda\sim
2M_{B_s}^2$  as an educative estimation. We find that the inclusion
of the real part contribution does not alter the order of magnitude
of the absorptive contribution. For example, when taking
$\phi_V=1$, we have,
\be\label{real}
{\cal BR}(B_s \rightarrow 2 D_s^*\rightarrow 2\gamma ) \sim
3.7\times 10^{-7}\ ,\,\,\,\, \Lambda=2M_{B_s}^2\ ,
\ee
to be compared with Eq.~(\ref{d*b}).

To conclude, we have re-evaluated the
$B_s\rightarrow 2\gamma$ decay  process through $D_s^+D_s^-$ and
$D_s^{*+}D_s^{*-}$ intermediate states' rescatterings. We have
estimated the absorptive part contribution to the decay branching
ratio in the effective Lagrangian approach, and also estimated the
real part contribution through a cut-off regulated dispersion
relation. We argue that the decay branching ratio will not exceed a
few times $10^{-7}$. Therefore if the furture experiments reveal a
large branching ratio of a few times $10^{-6}$ it must come from
new physics effects.

 {\it Acknowledgment}: The work of H.Z. is
supported in part by China National Natural Science Foundations.
%%%%%%%%%%%%%%%%%%%%%%%%%%%%%%%%%%%%%%%%%%%%%%%%%%%%%%%%%%%%%%%
% journal macros
%
\def\tp{these proceedings}
\def\ib#1,#2,#3{       {\em ibid.\/ }{\bf #1} (19#2) #3}
\def\ap#1,#2,#3{       {\em Ann.~Phys.~(NY)\/ }{\bf #1} (19#2) #3}
\def\appb#1,#2,#3{     {\em Acta Phys.\ Polon.\/ }{\bf B#1} (19#2) #3}
\def\cpc#1,#2,#3{      {\em Comput. Phys. Commun.\/ }{\bf #1} (19#2) #3}
\def\ijmp#1,#2,#3{     {\em Int.~J.~Mod.~Phys.\/ } {\bf A#1} (19#2) #3}
\def\mpl#1,#2,#3 {     {\em Mod.~Phys.~Lett.\/ } {\bf A#1} (19#2) #3}
\def\np#1,#2,#3{       {\em Nucl.~Phys.\/ }{\bf B#1} (19#2) #3}
\def\npps#1,#2,#3{     {\em Nucl.~Phys.~B (Proc.~Suppl.)\/ }
                             {\bf B#1} (19#2) #3}
\def\plb#1,#2,#3{      {\em Phys.~Lett.\/ }{\bf B#1} (19#2) #3}
\def\pr#1,#2,#3{       {\em Phys.~Rev.\/ }{\bf #1} (19#2) #3}
\def\prd#1,#2,#3{      {\em Phys.~Rev.\/ }{\bf D#1} (19#2) #3}
\def\prep#1,#2,#3{     {\em Phys.~Rep.\/ }{\bf #1} (19#2) #3}
\def\prl#1,#2,#3{      {\em Phys.~Rev.~Lett.\/ }{\bf #1} (19#2) #3}
\def\prog#1,#2,#3{     {\em Prog.~Theor.~Phys.\/ }{\bf #1} (19#2) #3}
\def\rmp#1,#2,#3{      {\em Rev.~Mod.~Phys.\/ }{\bf #1} (19#2) #3}
\def\sp#1,#2,#3{       {\em Sov.~Phys.-Usp.\/ }{\bf #1} (19#2) #3}
\def\zpc#1,#2,#3{      {\em Z. Phys.\/ }{\bf C#1} (19#2) #3}
%%%%%%%%%%%%%%%%%%%%%

\end{document}